\documentclass[referee]{raa}
\usepackage{graphicx,times}
\usepackage{natbib}
\usepackage{amssymb,amsmath}

\bibpunct{(}{)}{;}{a}{}{,}

\usepackage[a4paper=true,dvipdfm=true,pagebackref=true]{hyperref}
\hypersetup{pdftitle = The title of my PDF, pdfauthor = My name, pdfsubject= The subject, pdfkeywords = keyword1 keyword2 keyword3}
\hypersetup{colorlinks = true, linkcolor = blue, anchorcolor = red, citecolor = blue, filecolor = red, pagecolor = red, urlcolor = red}

\begin{document}

   \title{Correlation among extinction efficiency and other parameters in an aggregate dust model}

 \volnopage{ {\bf 2017} Vol.\ {\bf X} No. {\bf XX}, 000--000}
   \setcounter{page}{1}

   \author{Tanuj Kumar Dhar and Himadri Sekhar Das
   }

   \institute{Department of Physics, Assam University, Silchar 788011, India; {\it hsdas13@gmail.com}\\
\vs \no
   {\small Received 2017 Feb XX; accepted 2017 XXXX}
}

\abstract{We study the extinction properties of highly porous BCCA dust aggregates in a wide range of complex refractive indices ($1.4 \le n \le 2.0$, $0.001 \le k \le 1.0$) and wavelength ($0.11\mu m \le \lambda \le 3.4\mu m$). An attempt has been made for the first time to investigate the correlation among extinction efficiency ($Q_{ext}$), the composition of dust aggregates ($n, k$), the wavelength of radiation ($\lambda$) and size parameter of the monomers ($x$). If $k$ is fixed at any value between 0.001 and 1.0, $Q_{ext}$ increases with increase of $n$ from 1.4 to 2.0. $Q_{ext}$ and $n$ are correlated  via \emph{linear} regression when the cluster size is small whereas the correlation is \emph{quadratic} at moderate and higher sizes of the cluster. This feature is observed at all wavelengths (UV to optical to infrared). We also find that the variation of $Q_{ext}$ with $n$ is very small when $\lambda$ is high. When $n$ is fixed at any value between 1.4 and 2.0, it is observed that  $Q_{ext}$ and $k$ are correlated  via polynomial regression equation (of degree 1, 2, 3 or 4), where the degree of the equation depends on the cluster size, $n$ and $\lambda$. The correlation  is linear for small size and quadratic/cubic/quartic for moderate and higher sizes. We have also found that $Q_{ext}$ and $x$ are correlated via a polynomial regression (of degree 3,4 or 5) for all values of $n$. The degree of regression is found to be $n$ and $k$-dependent. The set of relations obtained from our work can be used to model interstellar extinction for dust aggregates in a wide range of wavelengths and complex refractive indices.
}

   \authorrunning{Dhar \& Das }            
   \titlerunning{Correlation among extinction efficiency and other parameters}  
   \maketitle

%
\section{Introduction}
The studies of cometary and interplanetary dust indicate that cosmic dust grains are likely to be fluffy, porous and
composites of many small grains fused together, due to dust-gas interactions, grain-grain collisions,
and various other processes \citep{b6,b3,b17}. Porous, composite aggregates are often modelled as cluster of small spheres (known as ``monomers"), agglomerated under various aggregation rules. Here grain aggregates are assumed to be fluffy sub structured collections of very small particles loosely attached to one another. Each particle is assumed to consist of a single material, such as silicates or carbon, as formed in the  various separate sources of cosmic dust. Extinction generally takes place whenever electromagnetic radiation propagates through a medium containing small particles.  The spectral dependence of extinction, or extinction curve, is a function of the structure, composition, and size distribution of the particles. The study of interstellar extinction provides us useful information for understanding the properties of the dust.

It is now well accepted from observation and laboratory analysis of interplanetary dust particles that cosmic dust grains are \textit{fluffy aggregates} or \textit{porous} with irregular shapes \citep{b1,b8,b3}. Using Discrete Dipole Approximation (DDA) technique, several investigators studied the extinction properties of the composite grains \citep{b17,b18,b16,b13,b14,b15}. \cite{b5} studied optical properties of composite grains as grain aggregates of amorphous carbon and astronomical silicates, using the Superposition transition matrix approach. Recently, \cite{b8a} studied the light scattering properties of aggregate particles in a wide range of complex refractive indices and wavelengths to investigate the correlation among different parameters e.g., the positive polarization maximum, the amplitude of the negative polarization, geometric albedo, refractive indices and wavelength. The simulations were performed using the Superposition T-matrix code with Ballistic Cluster–Cluster Aggregate (BCCA) particles of 128 monomers and Ballistic Aggre-
gates (BA) particles of 512 monomers.

The extinction efficiency of dust aggregates depends on aggregate size, composition and wavelength of incident radiation. The dependence of complex refractive index ($n, k)$ on $Q_{ext}$ was studied by many groups in past for spherical and irregular particles using different scattering theories (Mie theory, DDA approach, T-matrix theory etc.). But no correlation equations were reported earlier by any group. In this paper, we study the extinction properties of randomly oriented porous dust aggregates with a wide range of complex refractive indices and wavelength of incident radiation. An attempt has been made for the first time to investigate the correlation among extinction efficiency ($Q_{ext}$), complex refractive indices ($m = n + ik$), wavelength ($\lambda$) and the size parameter of monomer ($x$).

\section{Numerical computations}
We have constructed the aggregates using ballistic
aggregation procedure \citep{b9,b10} using two different
models of cluster growth. First via single-particle aggregation
and then through cluster-cluster aggregation. These aggregates are
built by random hitting and sticking particles together. The first
one is called Ballistic Particle-Cluster Aggregate (BPCA) when the
method allows only single particles to join the cluster of
particles. If the method allows clusters of particles to stick
together, the aggregate is called Ballistic Cluster-Cluster
Aggregate (BCCA). Actually, the BPCA clusters are more compact
than BCCA clusters \citep{b11}. The porosity of BPCA and
BCCA particles of 128 monomers has the values 0.90 and 0.94,
respectively. The fractal dimensions of BPCA and BCCA particles are given by $D \thickapprox 3$ and $\thickapprox 2$, respectively (Meakin 1984). A systematic explanation on dust aggregate model is
already discussed in our previous work \citep{b2}. It is to be noted that the
structure of these aggregates are similar to those of Interplanetary Dust Particles (IDP) collected in the stratosphere of Earth \citep{b1}.
It is also well understood from the laboratory diagnosis that the particle coagulation in the solar nebula grows under BCCA process \citep{b19}.

The general extinction $A_{\lambda}$ is given by \citep{b12}:
\begin{equation}
A_{\lambda} = -2.5 \log \left[{\frac {F(\lambda)}{F_0(\lambda)}}\right] = 1.086 \, N_d\, Q_{ext}\, \sigma _d,
\end{equation}
where $F(\lambda)$ and $F_0({\lambda})$ are the observed and expected fluxes, $N_d$ is the dust column density, $Q_{ext}$ is the extinction efficiency factor determined from Superposition T-matrix code, and $\sigma _d$ is the geometrical cross-section of a single particle.

The interstellar extinction curve (i.e. the variation of extinction with wavelength) is usually expressed by the ratio
A$_\lambda$/E(B-V) versus 1/$\lambda$. The extinction curve covers the wavelength range 0.11 to 3.4 $\mu m$. The entire range consists of UV (ultra violet), visible and IR (infrared) regions. The IR range  corresponds to near infrared i.e. 0.750 to 2.5 $\mu m$, the Visible range (0.38 to 0.76 $\mu m$) and the UV range (the last part of violet in visible spectrum to 0.11 $\mu m$).

The radius of an aggregate particle can be described by the radius of a sphere of equal volume given by $a_v = a_m$N$^{1/3}$, where N is the
number of monomers in the aggregate and $a_m$ is the monomer's radius of aggregates . We have found from literature survey that most of the work related to interstellar extinction considered a normal size range of 0.001 to 0.250 micron, with a size distribution (mainly MRN distribution) (Jones 1988, Whittet 2003, Vaidya et al. 2007, Das et al. 2010). They found an `optimum' for the range of the cluster size generally used. The above size range of the monomer is more or less capable of evaluating average observed interstellar extinction curve. If we consider N = 64 and $a_m$ in the range 0.001 to 0.065 micron (with a step size of 0.004 micron), then $a_v$ will be 0.004 to 0.26 micron. This size range is almost comparable to the size range used by other investigators.

We use JaSTA-2 (Second version of the Java Superposition T-matrix Application) \citep{b4a},  which is an upgraded version of JaSTA \citep{b4}, to execute our computations which is based on \cite{b7}'s Superposition T-matrix code. All versions of JaSTA are freely available to download from \verb"http://ausastro.in/jasta.html". The computations with the T-matrix code is fast and this technique gives rigorous solutions for randomly oriented ensembles of  spheres. It is to be noted that the results obtained from the Discrete Dipole Approximation (DDA) approach and the T-matrix approach are almost same. \cite{b5b} showed the results with aggregates using the DDA and the T-matrix code, and found almost same results with both the code. We perform the computations  with a wide range of complex refractive indices ($n$ = 1.4, 1.5, 1.6, 1.7, 1.8, 1.9, 2.0 and $k$ = 0.001, 0.01, 0.05, 0.1, 0.3, 0.5, 0.7, 1.0) and wavelengths (0.11, 0.12, 0.13, 0.16, 0.175, 0.185, 0.20, 0.207, 0.22, 0.23, 0.26, 0.30, 0.365, 0.40, 0.55, 0.6, 0.7, 0.8, 0.90 and 3.4 $\mu m$). In general, the range of $n$ and $k$ which is considered in our work almost covers the range of the complex refractive indices of silicate and carbon at different wavelengths. The numerical computation in the present work has been executed with BCCA cluster of 64 monomers.

We present the results for $a_m = 0.001\mu m, 0.017\mu m, 0.041\mu m$ and $0.065\mu m$ where $a_v$ is given by $0.004\mu m, 0.068\mu m, 0.16\mu m$ and $0.26\mu m$. The monomer size parameter ($x = 2 \pi a_m/\lambda$) is taken in a range from 0.01 to 1.6. This study is mainly concentrated on investigation of correlation among $Q_{ext}$, $(n, k)$, $\lambda$, and $x$.

\section{Results}
\subsection{Dependence on monomer size ($a_m$)}

\subsubsection{Correlation between $Q_{ext}$ and $n$}
We perform the computations with a wide range of complex refractive indices ($1.4 \le n \le   2.0$ and $0.001 \le k \le 1.0$)  and wavelength of incident radiation ($0.11 \le \lambda \le 3.4\mu m$). To study the dependence of $n$ and $k$ on the extinction efficiency ($Q_{ext}$), we can either plot $Q_{ext}$ versus $k$ by keeping $n$ fixed or plot $Q_{ext}$ versus $n$ by keeping $k$ fixed. We first show the results for moderate size of the cluster ,i.e., at $a_v = 0.16\mu m$. The results at other three sizes are also presented thereafter.

\begin{figure*}
\begin{center}
\vspace{12 cm}
\includegraphics[]{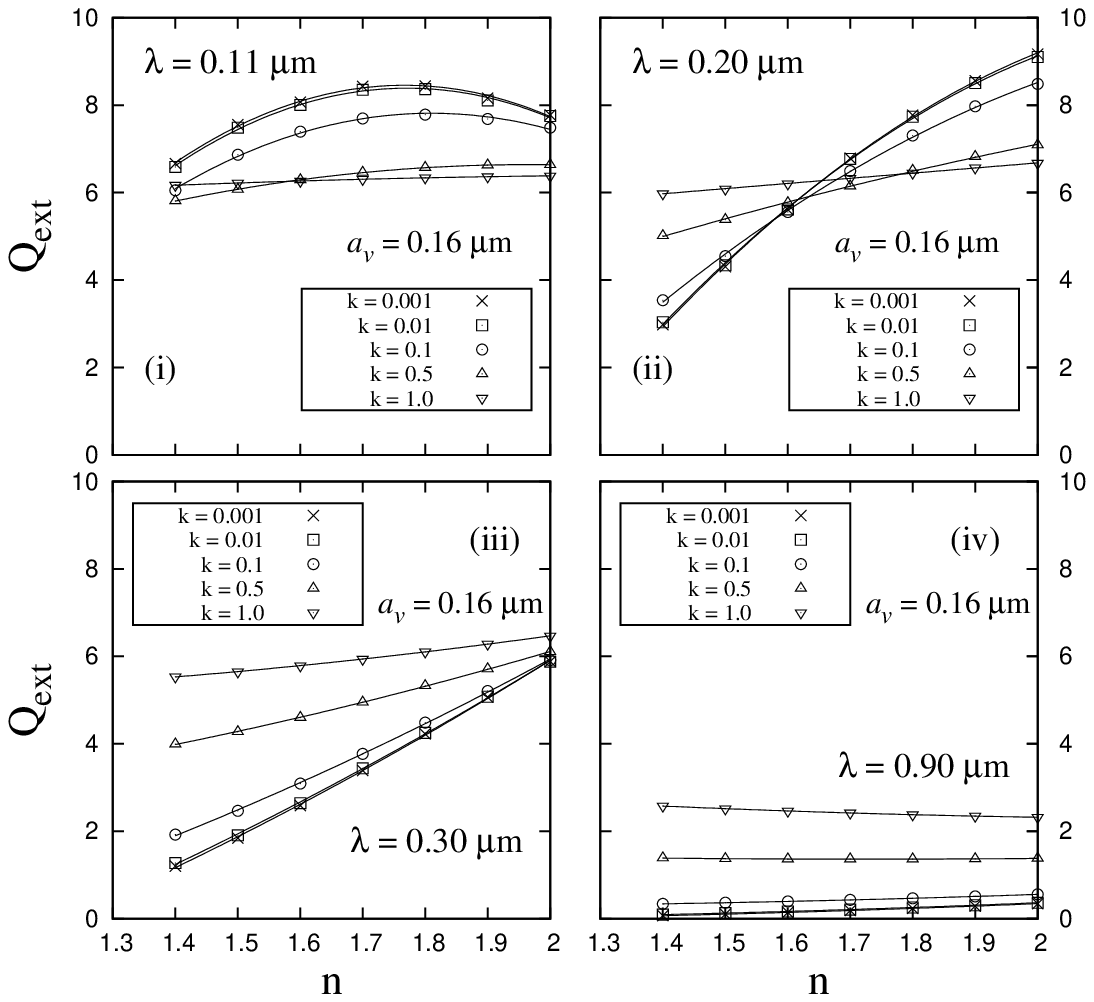}
\vspace{-9cm}
\caption{Extinction efficiency ($Q_{ext}$) is plotted against real part of the refractive index ($n$) for $k$ = 0.001, 0.01, 0.1, 0.5 and 1.0 at $a_v = 0.16\mu m$. The best fit curves correspond to  \emph{quadratic regression} of the form $Q_{ext} = A_kn^2 + B_kn + C_k$ for wavelengths (i) 0.11 $\mu m$, (ii) 0.20 $\mu m$, (iii) 0.30 $\mu m$ and (iv) 0.90 $\mu m$.}
\label{Fig1}
\end{center}
\end{figure*}

We plot $Q_{ext}$ versus $n$ at $a_v = 0.16\mu m$, for $k$ = 0.001, 0.01, 0.1, 0.5 and 1.0 respectively (although we have executed the code with all eight values of $k$ mentioned above), in a single frame, for $\lambda = 0.11, 0.20, 0.30$ and $0.90\mu m$, which is shown in Fig.\ref{Fig1}. It is to be noted that the plots are shown for four wavelengths although the computations have been performed for all the wavelengths.

It is observed from Fig.\ref{Fig1} that if $k$ is fixed at any value between 0.001 and 1.0, $Q_{ext}$ increases with increase of $n$ from 1.4 to 2.0 at all wavelengths. But a small decrease in $Q_{ext}$ is also noticed at $\lambda = 0.11\mu m$ when $n > 1.7$ and $k$ is low. The variation of $Q_{ext}$ with $n$ is small when $k \ge 0.5$. The value of $Q_{ext}$ is small when $\lambda$ is large, i.e. $Q_{ext}$ decreases when size parameter of the monomer ($x = 2 \pi a_m/\lambda$) decreases. We have also investigated  that the variation of $Q_{ext}$ with $n$ is very small when $\lambda \ge 0.70\mu m$.

We have found that $Q_{ext}$ and $n$ can be fitted by \emph{quadratic regression} where \emph{coefficient of determination}\footnote{\emph{The coefficient of determination} is a key output of regression analysis which is interpreted as the proportion of the variance in the dependent variable which ranges from 0 to 1. A higher coefficient is an indicator of a better goodness of fit for the observations.} ($R^2$) for each equation is $\approx$ 0.99. The best fit equation is given by
\begin{subequations}\label{eq:2}
\begin{equation}\tag{2}
\label{1a}
Q_{ext} = A_k n^2 + B_k n + C_k~~~, ~~~~1.4 \le n \le 2.0, ~~~~~~ 0.001 \le k \le 1,
\end{equation}
where, $A_k, B_k$ and $C_k$ are $k$-dependent coefficients of equation (2).

The coefficients obtained for different values of $k$ (only five values of $k$ are shown) are depicted in Table-\ref{table:1}. If we plot coefficients $A_k, B_k$ and $C_k$  versus $k$ (where $k$ = 0.001, 0.01, 0.05, 0.1, 0.3, 0.5, 0.7, 1.0), we find that the best fit curves correspond to \emph{cubic regression}, which have $R^2$ $\approx$ 0.99. We do  not show any figures in this case.

The coefficients are correlated with $k$ by the relations:
\begin{align}
~~~~~~~             \tag{2a}        A_k =& D_1 k^3 + D_2 k^2 + D_3 k + D_4~~,\\
~~~~~~~             \tag{2b}        B_k =& E_1 k^3 + E_2 k^2 + E_3 k + E_4~~,\\
~~~~~~~             \tag{2c}        C_k =& F_1 k^3 + F_2 k^2 + F_3 k + F_4~~,
 \end{align}
\end{subequations}
All coefficients of equation 2(a-c)  are shown in Table-\ref{table:2}. Thus knowing the coefficients, the extinction efficiency ($Q_{ext}$) can be calculated for any value of $n$ and $k$ from the equation (2).

\begin{table*}
\begin{center}
\caption{Co-efficients of equation (2) are shown for $\lambda = 0.11, 0.20, 0.30$ and $0.90\mu m$.}
\begin{tabular}{|l|l|r|r|r|}
  \hline
$\lambda$ & $k$	 &	$A_k$	&	$B_k$	&	$C_k$\\
\hline	
0.11 $\mu m$& 0.001	&	$-$13.245	&	46.876	&	$-$33.026	\\
    & 0.01	&	$-$12.937	&	45.862	&	$-$32.263	\\
    & 0.1	&	$-$10.123	&	36.537	&	$-$25.149	\\
    & 0.5	&	$-$2.651	&	10.490	&	$-$3.731	\\
    &  1.0	&	$-$0.346	&	1.555	&	4.661	\\
	\hline						
0.20 $\mu m$ & 0.001	&	$-$7.815	&	36.995	&	$-$33.540	\\
     & 0.01	&	$-$7.572	&	35.969	&	$-$32.527	\\
     & 0.1	&	$-$5.439	&	26.899	&	$-$23.511	\\
     & 0.5	&	$-$0.969	&	6.779	&	$-$2.590	\\
     & 1.0	&	0.010	   &	1.141	&	4.359	\\
\hline			
0.30 $\mu m$ & 0.001	&	1.651	&	2.290	&	$-$5.271	\\
     & 0.01	&	1.647	&	2.191	&	$-$5.054	\\
     & 0.1	&	1.589	&	1.335	&	$-$3.095	\\
     & 0.5	&	3.527	&	$-$0.999	&	3.939	\\
     & 1.0	&	0.681	&	$-$0.749	&	5.243	\\
\hline			
0.90 $\mu m$ & 0.001	&	0.217	&	$-$0.278	&	0.032	\\
     & 0.01	&	0.217	&	$-$0.285	&	0.068	\\
     & 0.1	&	0.213	&	$-$0.363	&	0.429	\\
     & 0.5	&	0.230	&	$-$0.794	&	2.046	\\
     & 1.0	&	0.328	&	$-$1.543	&	4.092\\
\hline
\end{tabular}
\label{table:1}
\end{center}
\end{table*}


\begin{table*}
\begin{center}
\caption{Co-efficients of equation 2(a-c) at $\lambda = 0.11, 0.20, 0.30$ and 0.90 $\mu m$.}
\begin{tabular}{|l|l|r|r|r|r|}
  \hline
	&		&	Coeff-1	&	Coeff-2	&	Coeff-3	&	Coeff-4	\\
\hline
$\lambda = 0.11\mu m$	&	$A_k$	&	10.161	&	$-$31.889	&	34.662	 &	$-$13.280	\\
	&	$B_k$	&	$-$26.328	&	94.623	&	$-$113.730	&	46.990	\\
	&	$C_k$	&	11.321	&	$-$58.956	&	85.407	&	$-$33.111	\\
\hline											
$\lambda = 0.20\mu m$	&	$A_k$	&	15.474	&	$-$35.001	&	27.379	 &	$-$7.842	\\
	&	$B_k$	&	$-$60.269	&	139.790	&	$-$115.490	&	37.110	\\
	&	$C_k$	&	55.593	&	$-$131.620	&	114.040	&	$-$33.654	\\
\hline											
$\lambda = 0.30\mu m$	&	$A_k$	&	1.586	&	$-$2.127	&	$-$0.431	 &	1.652	\\
	&	$B_k$	&	$-$8.473	&	16.673	&	$-$11.251	&	2.302	\\
	&	$C_k$	&	10.310	&	$-$24.081	&	24.309	&	$-$5.295	\\
\hline											
$\lambda = 0.90\mu m$	&	$A_k$	&	0.001	&	0.171	&	$-$0.060	 &	0.217	\\
	&	$B_k$	&	$-$0.024	&	$-$0.430	&	$-$0.812	&	$-$0.277	 \\
	&	$C_k$	&	0.002	&	0.053	&	4.009	&	0.028	\\
\hline											
\end{tabular}
\label{table:2}
\end{center}
\end{table*}

\begin{figure*}
\begin{center}
\vspace{12 cm}
\includegraphics[]{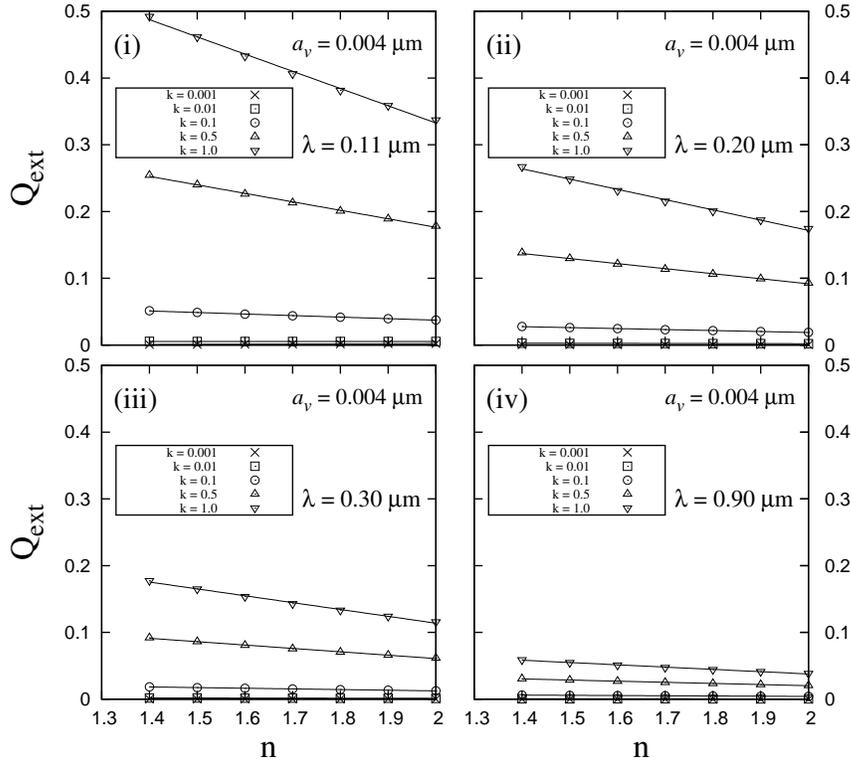}
\vspace{-9cm}
\caption{Extinction efficiency ($Q_{ext}$) is plotted against real part of the refractive index ($n$) for $k$ = 0.001, 0.01, 0.1, 0.5 and 1.0 at $a_v = 0.004\mu m$. The best fit curves correspond to  \emph{linear regression}  for wavelengths (i) 0.11 $\mu m$, (ii) 0.20 $\mu m$, (iii) 0.30 $\mu m$ and (iv) 0.90 $\mu m$.}
\label{Fig2}
\end{center}
\end{figure*}

\begin{figure*}
\begin{center}
\vspace{12 cm}
\includegraphics[]{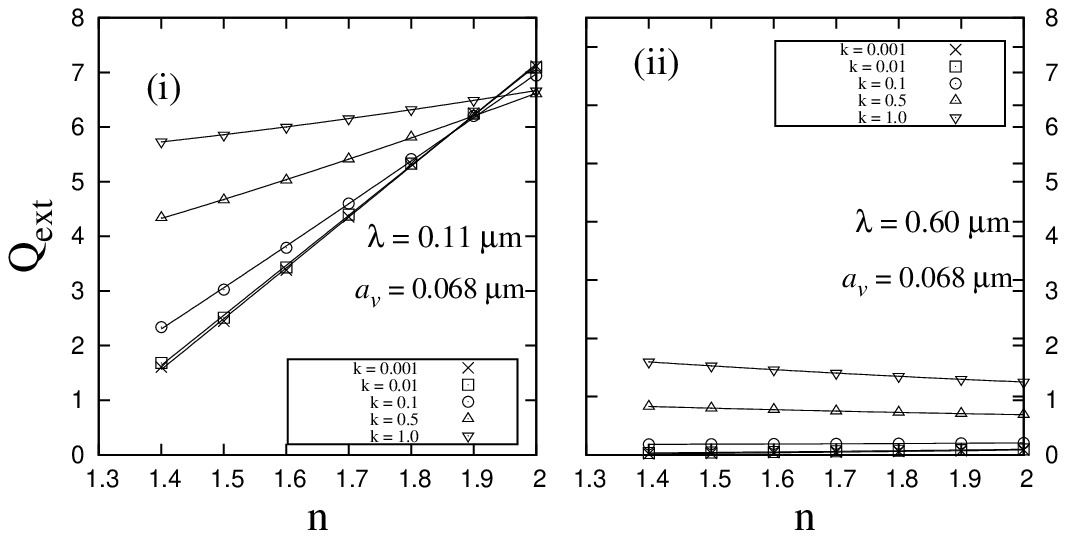}
\vspace{-15cm}
\caption{Extinction efficiency ($Q_{ext}$) is plotted against real  part of the refractive index ($n$) for $k$ = 0.001, 0.01, 0.1, 0.5 and 1.0 at $a_v = 0.068\mu m$. The best fit curves correspond to  \emph{quadratic regression} for wavelengths (i) 0.11 $\mu m$ and (ii) 0.60 $\mu m$.}
\label{Fig3}
\end{center}
\end{figure*}

\begin{figure*}
\begin{center}
\vspace{12 cm}
\includegraphics[]{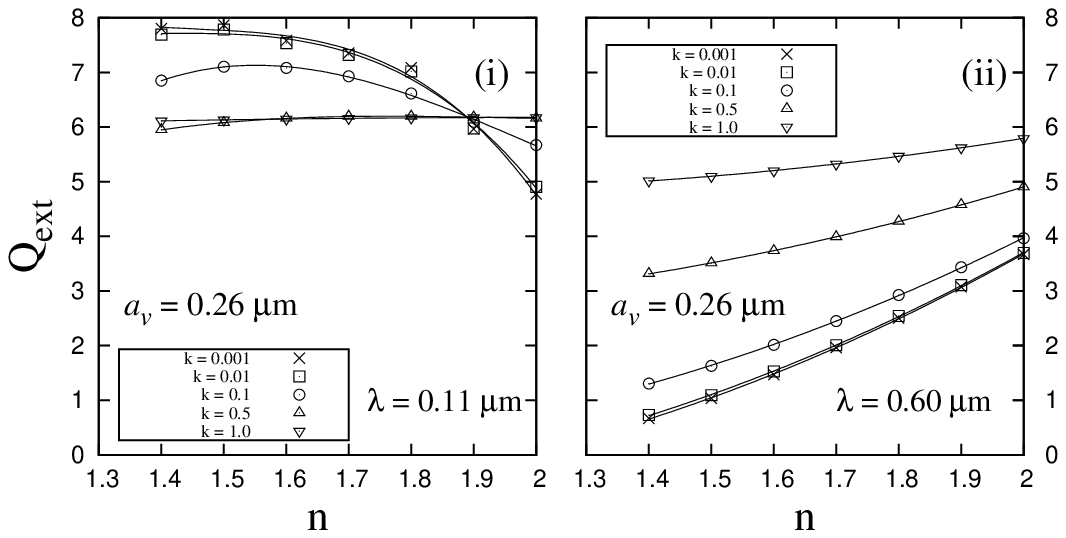}
\vspace{-15cm}
\caption{Extinction efficiency ($Q_{ext}$) is plotted against real  part of the refractive index ($n$) for $k$ = 0.001, 0.01, 0.1, 0.5 and 1.0 at $a_v = 0.26\mu m$. The best fit curves correspond to  \emph{cubic regression} at (i) $\lambda = 0.11\mu m$ and \emph{quadratic regression} at (ii) $\lambda = 0.60\mu m$.}
\label{Fig4}
\end{center}
\end{figure*}

In Fig.\ref{Fig2}, we report the results for $a_v = 0.004\mu m$ at $\lambda = 0.11, 0.20, 0.30$ and 0.90 $\mu m$. A strong \emph{linear} correlation between $Q_{ext}$ and $n$ is seen at this size for all wavelengths from 0.11 to 3.4 $\mu m$. In Fig.\ref{Fig3}, we plot $Q_{ext}$ versus $n$ for $a_v = 0.068\mu m$ at $\lambda = 0.11\mu m$ and 0.60 $\mu m$. We have found that the nature is \emph{quadratic} at all wavelengths from 0.11 to 3.4 $\mu m$. Finally, we show the results for $a_v = 0.26\mu m$ at $\lambda = 0.11\mu m$ and 0.60 $\mu m$, shown in Fig.\ref{Fig4}. We have noticed that the $Q_{ext}$ and $n$ is correlated via a \emph{cubic} regression at 0.11 $\mu m$ whereas the dependence is \emph{quadratic} at other higher wavelengths. We do not show any equation or table in the above three cases. In summary, we can conclude that the correlation between $Q_{ext}$ and $n$ is \emph{linear} when the cluster size is small whereas the correlation is \emph{quadratic} at moderate and higher sizes of cluster.


\subsubsection{Correlation between $Q_{ext}$ and $k$}

\begin{figure*}
\begin{center}
\vspace{12 cm}
\includegraphics[]{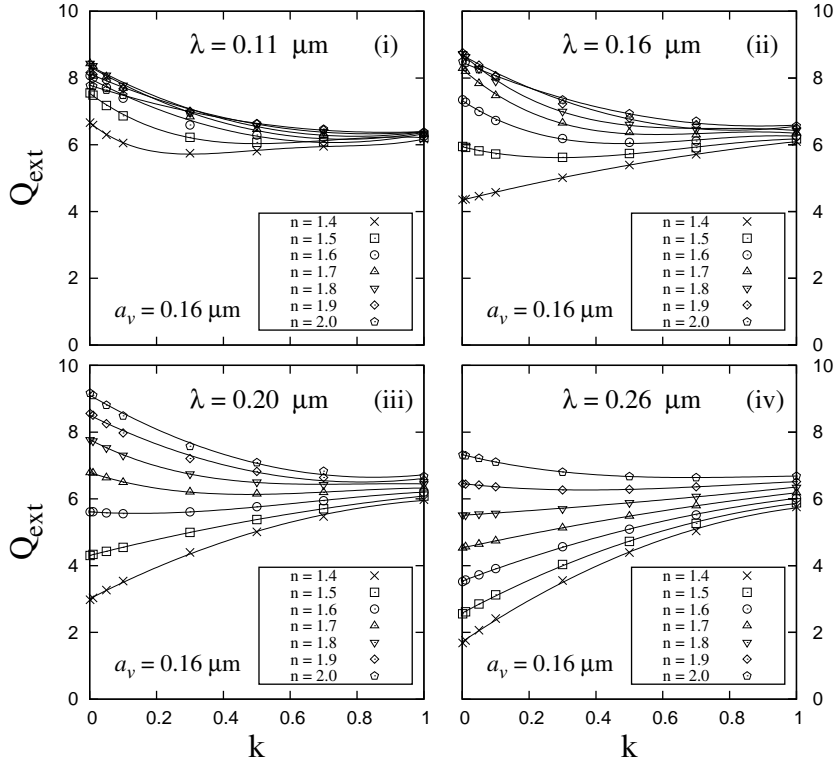}
\vspace{-9cm}
\caption{Extinction efficiency ($Q_{ext}$) is plotted against imaginary  part of the refractive index ($k$) at $\lambda$ = 0.11, 0.16, 0.20 and  0.26 $\mu m$ at $a_v = 0.16\mu m$. The best fit curves correspond to polynomial regression equation where the degree of equation depends on the value of $n$ and $\lambda$ (please see \emph{Section 3.2} for details).}
\label{Fig5}
\end{center}
\end{figure*}

We now plot $Q_{ext}$ versus $k$ for $n$ = 1.4, 1.5, 1.6, 1.7, 1.8, 1.9 and 2.0 respectively, at $a_v = 0.16\mu m$ where $\lambda$ is taken in between 0.11 $\mu m$ and 3.4 $\mu m$. When $\lambda$ is between 0.11 $\mu m$ and 0.26 $\mu m$, we have found that $Q_{ext}$ and $k$ can be fitted via a polynomial regression equation where the degree of equation depends on the value of $n$ and $\lambda$. In Fig.\ref{Fig5} we show the plots for $\lambda = 0.11, 0.16, 0.20$ and $0.26\mu m$. At $\lambda = 0.11\mu m$, we find that $Q_{ext}$ and $k$ are correlated by (i) \emph{quartic} regression when $n = 1.4$, (ii) \emph{cubic} regression when $n = 1.5$ and (iii) \emph{quadratic} regression when $n$ = 1.6, 1.7, 1.8, 1.9 \& 2.0. Further at $\lambda = 0.16\mu m$, the correlation is \emph{cubic} when $n$ = 1.5, 1.6, 1.7 \& 1.8 and \emph{quadratic} when $n$ = 1.4, 1.9 \& 2.0. The correlation at $\lambda = 0.20\mu m$ is \emph{cubic} when $n$ = 1.6, 1.7 \& 1.8 and \emph{quadratic} when $n$ = 1.4, 1.5, 1.9 \& 2.0. We also note that the correlation at 0.26 $\mu m$ is \emph{quadratic} when $n$ = 1.4, 1.5, 1.6, 1.7 \& 1.8 and \emph{cubic} when $n$ = 1.9 \& 2.0. At low values of $n$, $Q_{ext}$ increases with increase of $k$ whereas the trend is exactly opposite when $n$ is high ,e.g., at $\lambda = 0.20\mu m$, $Q_{ext}$ increases with $k$ when $n \le 1.6$, but it decreases when $n \ge 1.7$. The vertical range of $Q_{ext}$ in the plot also decreases when $k$ increases. This range is maximum at $k = 0.001$  and minimum at $k = 1.0$.

\begin{figure*}
\begin{center}
\vspace{12 cm}
\includegraphics[]{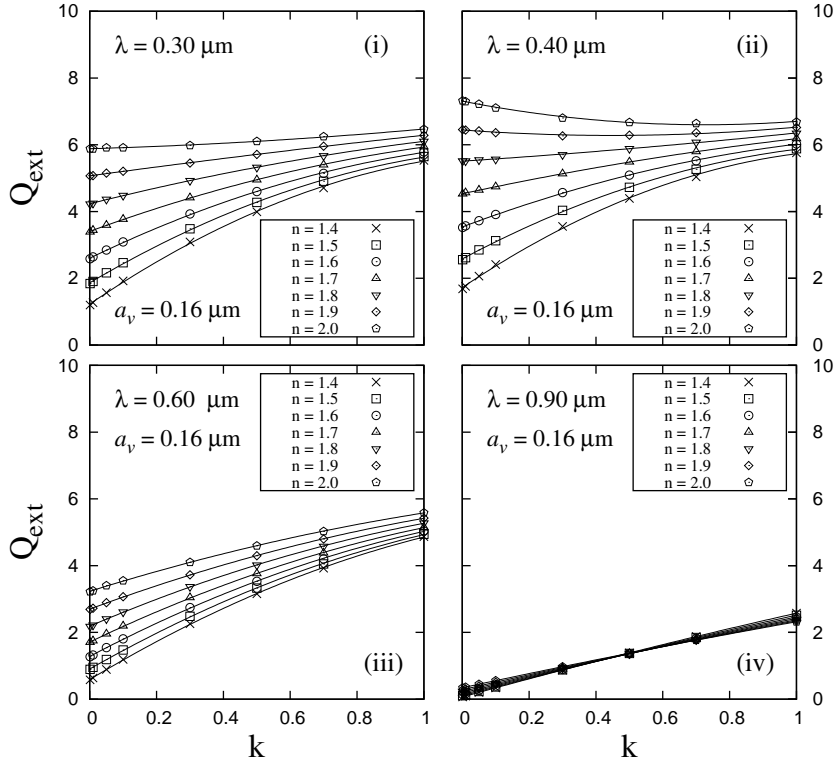}
\vspace{-9cm}
\caption{Extinction efficiency ($Q_{ext}$) is plotted against imaginary part of the  refractive index ($k$) for $n$ = 1.4, 1.5, 1.6, 1.7, 1.8, 1.9 and 2.0 at $a_v = 0.16\mu m$. The best fit curves correspond to  \emph{quadratic regression} of the form $Q_{ext} = A'_n k^2 + B'_n k + C'_n$ for wavelengths (i) 0.30 $\mu m$, (ii) 0.40 $\mu m$, (iii) 0.60 $\mu m$ and (iv) 0.90 $\mu m$.}
\label{Fig6}
\end{center}
\end{figure*}

In Fig.\ref{Fig6}, we show the plots for $0.30 \le \lambda \le 3.4\mu m$ and we have found that the fit is \emph{quadratic} for all values of $n$. An increase in $Q_{ext}$ with $k$ is noticed at almost all wavelengths. The vertical range of $Q_{ext}$ also decreases when $k$ increases. Further, the plot of $Q_{ext}$ with $k$ is same at all values of $n$ when $\lambda$ is high (please see Fig.\ref{Fig6}(iv)).

The best fit equation in the wavelength range 0.30 $\mu m$ to 3.4 $\mu m$ is given by
\begin{subequations}\label{eq:3}
\begin{equation}\tag{3}
\label{2a}
Q_{ext} = A_n k^2 + B_n k + C_n~~, ~~1.4 \le n \le 2.0, ~~ 0.001 \le k \le 1
\end{equation}
where, $A_n, B_n$ and $C_n$ are $n$-dependent coefficients of equation (3).

The coefficients obtained for different values of $k$ are shown in Table-\ref{table:3}. If we plot coefficients $A_n, B_n$ and $C_n$  versus $n$ (figures are not shown), we note that the best fit curves correspond to \emph{quadratic regression}, which have $R^2$ $\approx$ 0.99.

Thus coefficients are given by
\begin{align}
~~~~~~~             \tag{3a}        A_n =& D'_1 k^2 + D'_2 k + D'_3~~,\\
~~~~~~~             \tag{3b}        B_n =& E'_1 k^2 + E'_2 k + E'_3~~,\\
~~~~~~~             \tag{3c}        C_n =& F'_1 k^2 + F'_2 k + F'_3~~,
 \end{align}
\end{subequations}
All coefficients of equation 3(a-c)  are shown in Table-\ref{table:4}. Thus, knowing the coefficients of equation 3(a-c), the extinction efficiency ($Q_{ext}$) can be also estimated for any value of $n$ and $k$ from the equation (3).

\begin{table*}
\begin{center}
\caption{Co-efficients of equation (3) at $\lambda = 0.30, 0.40, 0.60$ and 0.90 $\mu m$.}
\begin{tabular}{|l|l|l|l|l|}
  \hline
$\lambda$	&	$n$	&	$A_n$	&	$B_n$	&	$C_n$	\\
\hline
0.30 $\mu m$	&	1.4	&	-2.579	&	6.890	&	1.195	\\
	&	1.5	&	-2.107	&	5.844	&	1.896	\\
	&	1.6	&	-1.630	&	4.774	&	2.632	\\
	&	1.7	&	-1.148	&	3.678	&	3.401	\\
	&	1.8	&	-0.661	&	2.556	&	4.204	\\
	&	1.9	&	-0.169	&	1.410	&	5.041	\\
	&	2.0	&	0.327	&	0.238	&	5.911	\\
		\hline							
0.40 $\mu m$	&	1.4	    &	-1.769	&	6.047	&	0.579	\\
	&	1.5	&	-1.655	&	5.675	&	0.912	\\
	&	1.6	&	-1.519	&	5.256	&	1.290	\\
	&	1.7	&	-1.362	&	4.789	&	1.712	\\
	&	1.8	&	-1.184	&	4.274	&	2.178	\\
	&	1.9	&	-0.984	&	3.712	&	2.688	\\
	&	2.0	&	-0.763	&	3.102	&	3.242	\\
		\hline							
0.60 $\mu m$	&	1.4	&	-0.727	&	4.222	&	0.198	\\
	&	1.5	&	-0.704	&	4.081	&	0.307	\\
	&	1.6	&	-0.676	&	3.931	&	0.433	\\
	&	1.7	&	-0.642	&	3.771	&	0.574	\\
	&	1.8	&	-0.602	&	3.602	&	0.731	\\
	&	1.9	&	-0.556	&	3.423	&	0.905	\\
	&	2.0	&	-0.505	&	3.234	&	1.094	\\
		\hline							
0.90 $\mu m$	&	1.4	&	-0.260	&	2.771	&	0.064	\\
	&	1.5	&	-0.257	&	2.674	&	0.100	\\
	&	1.6	&	-0.250	&	2.575	&	0.139	\\
	&	1.7	&	-0.240	&	2.475	&	0.183	\\
	&	1.8	&	-0.227	&	2.374	&	0.231	\\
	&	1.9	&	-0.210	&	2.272	&	0.284	\\
	&	2.0	&	-0.190	&	2.169	&	0.341	\\
		\hline							
\end{tabular}
\label{table:3}
\end{center}
\end{table*}

\begin{table*}
\begin{center}
\caption{Co-efficients of equation 3(a-c) at $\lambda = 0.30, 0.40, 0.60$ and 0.90 $\mu m$.}
\begin{tabular}{|c|c|r|r|r|}
  \hline
$\lambda$	&		&	Coeff-1	&	Coeff-2	&	Coeff-3	\\
				\hline					
0.30 $\mu m$	&	$A_n$	&	0.243	&	4.016	&	-8.679	\\
	&	$B_n$	&	-1.269	&	-6.771	&	18.857	\\
	&	$C_n$	&	1.682	&	2.142	&	-5.101	\\
	\hline								
0.40 $\mu m$	&	$A_n$	&	1.069	&	-1.956	&	-1.126	\\
	&	$B_n$	&	-2.385	&	3.201	&	6.241	\\
	&	$C_n$	&	2.202	&	-3.048	&	0.530	\\
	\hline								
0.60 $\mu m$	&	$A_n$	&	0.290	&	-0.614	&	-0.434	\\
	&	$B_n$	&	-0.481	&	-0.010	&	5.179	\\
	&	$C_n$	&	0.796	&	-1.214	&	0.337	\\
	\hline								
0.90 $\mu m$	&	$A_n$	&	0.171	&	-0.466	&	0.056	\\
	&	$B_n$	&	-0.060	&	-0.799	&	4.008	\\
	&	$C_n$	&	0.217	&	-0.278	&	0.028	\\
		\hline								
\end{tabular}
\label{table:4}
\end{center}
\end{table*}

\begin{figure*}
\begin{center}
\vspace{12 cm}
\includegraphics[]{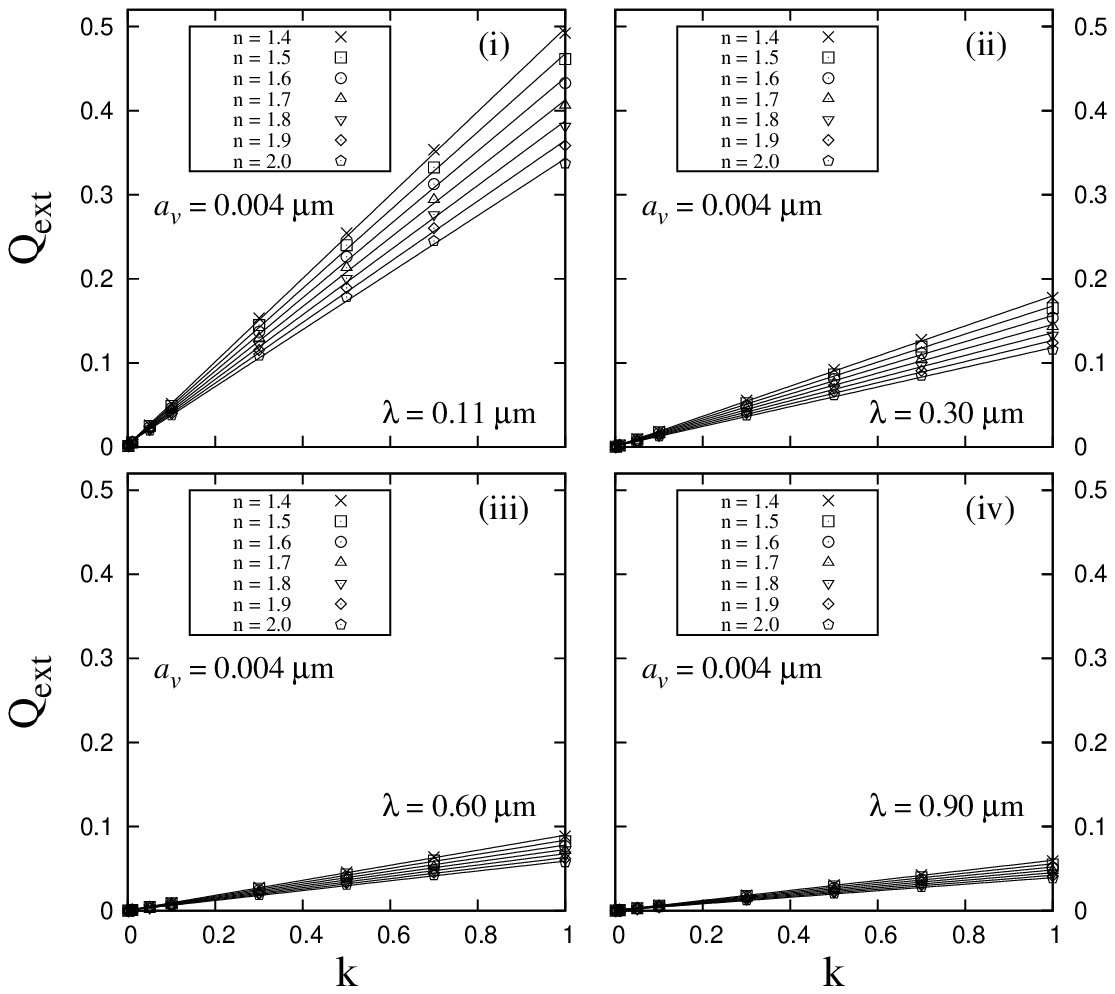}
\vspace{-9cm}
\caption{Extinction efficiency ($Q_{ext}$) is plotted against imaginary part of the  refractive index ($k$) for $n$ = 1.4, 1.5, 1.6, 1.7, 1.8, 1.9 and 2.0 at $a_v = 0.004\mu m$. The best fit curves correspond to  \emph{linear regression} for wavelengths (i) 0.11 $\mu m$, (ii) 0.30 $\mu m$, (iii) 0.60 $\mu m$ and (iv) 0.90 $\mu m$.}
\label{Fig7}
\end{center}
\end{figure*}

\begin{figure*}
\begin{center}
\vspace{12 cm}
\includegraphics[]{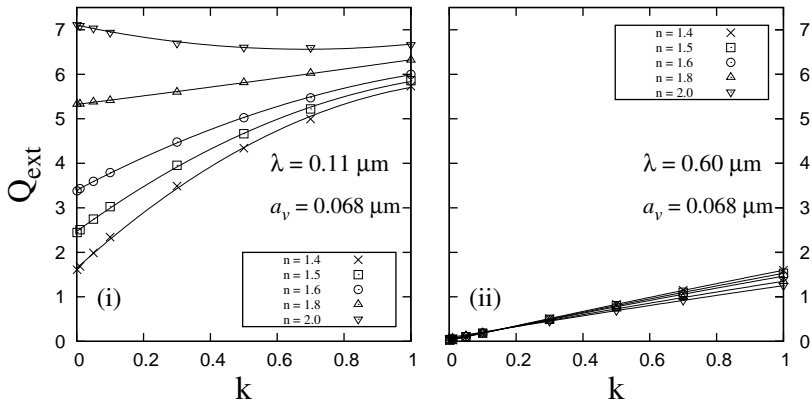}
\vspace{-15cm}
\caption{Extinction efficiency ($Q_{ext}$) is plotted against imaginary  part of the refractive index ($k$) for $n$ = 1.4, 1.5, 1.6, 1.8 and 2.0 at $a_v = 0.068\mu m$. The best fit curves correspond to  \emph{quadratic regression}  for wavelengths (i) 0.11 $\mu m$ and (ii) 0.60 $\mu m$.}
\label{Fig8}
\end{center}
\end{figure*}

\begin{figure*}
\begin{center}
\vspace{12 cm}
\includegraphics[]{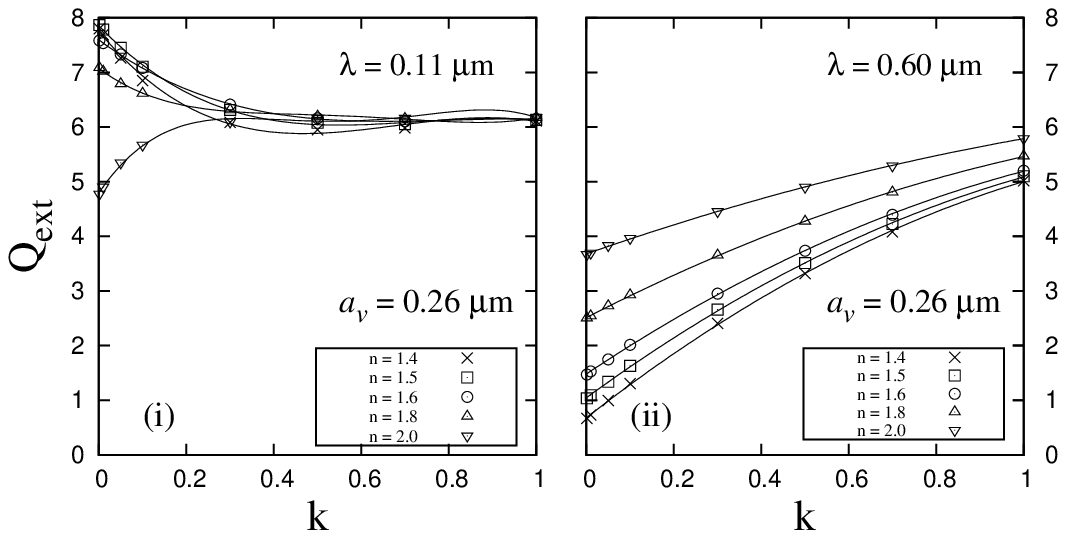}
\vspace{-15cm}
\caption{Extinction efficiency ($Q_{ext}$) is plotted against imaginary  part of the refractive index ($k$) for $n$ = 1.4, 1.5, 1.6, 1.8 and 2.0 at $a_v = 0.26\mu m$. $Q_{ext}$ and $k$ are correlated via polynomial regression equations, where the degree of regression is found to be wavelength dependent: (i) at $\lambda = 0.11\mu m$, the correlation is \emph{cubic} for $n = 1.4, 1.5, 1.6, 1.7$ and $2.0$, and \emph{quartic} at $n = 1.8$ and $1.9$, and (ii) at $\lambda = 0.60\mu m$, the correlation is \emph{quadratic} at all values of $n$.}
\label{Fig9}
\end{center}
\end{figure*}

In Fig.\ref{Fig7}, we plot $Q_{ext}$ against $k$ for $a_v = 0.004\mu m$ at $\lambda = 0.11, 0.30, 0.60$ and 0.90 $\mu m$, although the computations have been performed for wide range of wavelengths from 0.11 to 3.4 $\mu m$. We observe the \emph{linear} dependence at this size for all wavelengths. This linear nature becomes \emph{quadratic} when the size of cluster is $a_v = 0.068\mu m$. Fig.\ref{Fig8} shows the results for $a_v = 0.068\mu m$ at $\lambda$ = 0.11 $\mu m$ and 0.60 $\mu m$. In Fig.\ref{Fig9}, the results obtained for $a_v = 0.26\mu m$ are plotted. In this case, the nature of dependence looks similar with $a_v = 0.16\mu m$. $Q_{ext}$ and $k$ are correlated  via polynomial regression equation (of degree 2, 3 or 4) where the degree of equation depends on the real  part of the refractive index ($n$) at ($\lambda = 0.11\mu m$) (please see caption of Fig.\ref{Fig9}). The nature is quadratic when $\lambda > 0.11\mu m$. In summary, we can conclude that the dependence of $Q_{ext}$ on $k$ depends on the cluster size. The correlation is linear for small size and quadratic/cubic for moderate and higher sizes.

It is important to mention that the real part of the complex index of refraction ($n$) controls the effective phase speed of electromagnetic waves propagating through the medium, while the imaginary part $k$ describes the rate of absorption of the wave. In any material, $n$ and $k$ are not free to vary independently of one another but rather are tightly coupled to one another via the so-called Kramer-Kronig relations. Therefore, the results presented above are quite expectable.

\subsection{Dependence on wavelength of radiation ($\lambda$)}
\begin{figure*}
\begin{center}
\vspace{12 cm}
\hspace{-3 cm}
\includegraphics[]{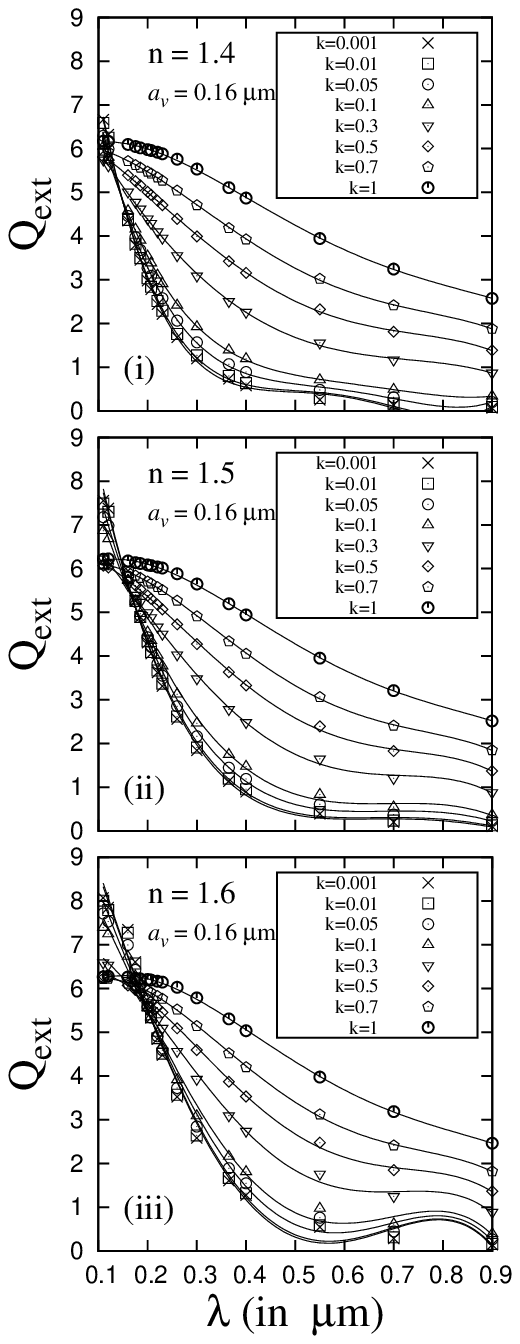}
\vspace{-6cm}
\caption{Extinction efficiency ($Q_{ext}$) is plotted against wavelength ($\lambda$) for $k$ = 0.001, 0.01, 0.05, 0.1, 0.3, 0.5, 0.7 and 1.0 at at $a_v = 0.16\mu m$ when (i) $n = 1.4$, (ii) $n= 1.5$ and (iii) $n= 1.6$. The best fit curves correspond to \emph{quartic regression}. }
\label{Fig10}
\end{center}
\end{figure*}

\begin{figure*}
\begin{center}
\vspace{12 cm}
\hspace{-2 cm}
\includegraphics[]{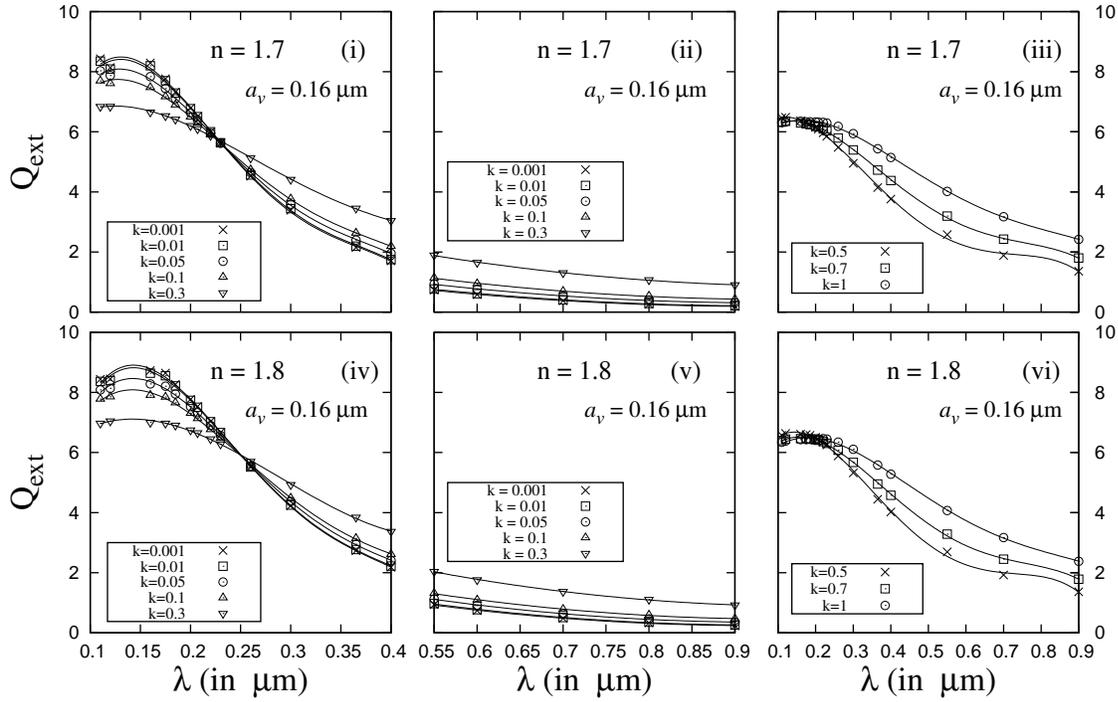}
\vspace{-9cm}
\caption{Extinction efficiency ($Q_{ext}$) is plotted against wavelength of incident radiation ($\lambda$) at $a_v = 0.16\mu m$. The left panel (top and bottom) shows the plot for imaginary  part of the refractive indices ($k$) = 0.001, 0.01, 0.05, 0.1 \& 0.3 in the wavelength range 0.11 $\mu m$ to 0.40 $\mu m$, the middle panel (top and bottom) is for $k$ = 0.001, 0.01, 0.05, 0.1 \& 0.3  in the wavelength range 0.55 $\mu m$ to 0.90 $\mu m$ and the right panel (top and bottom) is for $k$ = 0.5, 0.7 \& 1 in the wavelength range 0.11 $\mu m$ to 0.90 $\mu m$ . The best fit corresponds to \emph{quartic} regression (left panel), \emph{quadratic} regression (middle panel) and \emph{quartic} regression (right panel).  The real  part of the refractive index ($n$) is fixed at 1.7 and 1.8.}
\label{Fig11}
\end{center}
\end{figure*}

\begin{figure*}
\begin{center}
\vspace{12 cm}
\hspace{-2 cm}
\includegraphics[]{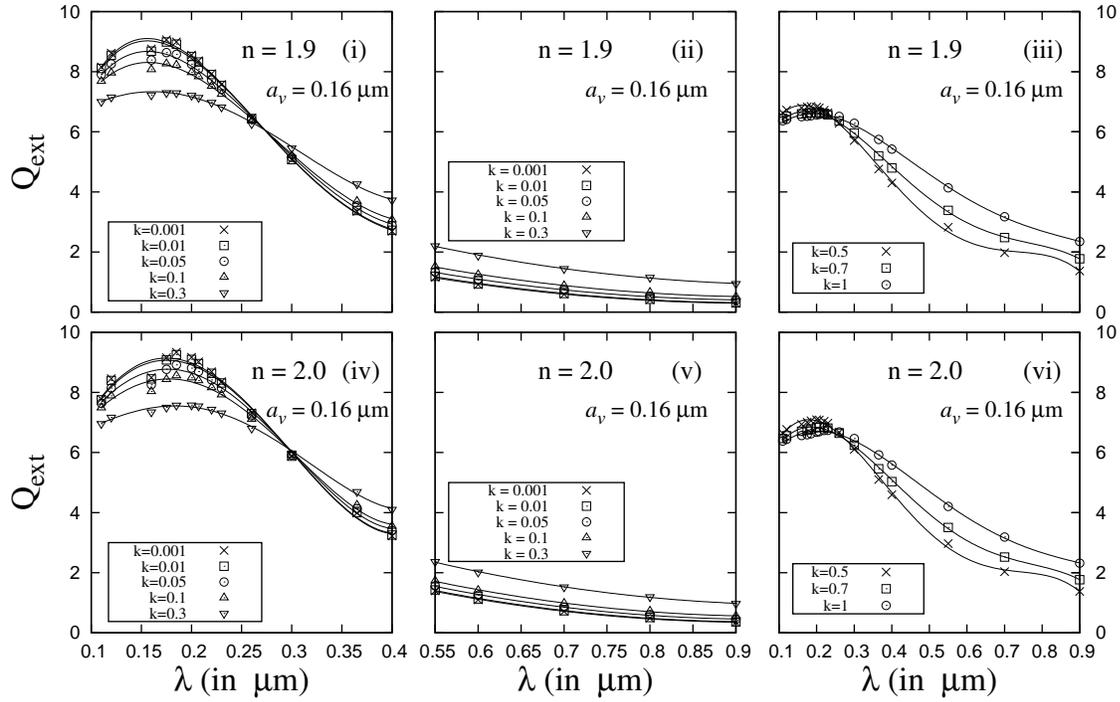}
\vspace{-9cm}
\caption{Same as Fig.\ref{Fig11} but with $n$ = 1.9 and 2.0.}
\label{Fig12}
\end{center}
\end{figure*}

We now study the dependence of $Q_{ext}$ on $\lambda$ for a particular set of $(n, k)$ in case of $a_v = 0.16\mu m$ only. We observe the following results:
\begin{enumerate}
  \item[(i)] For $n$ = 1.4, 1.5 and 1.6, $Q_{ext}$ versus $\lambda$ can be fitted via a \emph{quartic} regression for $k$ = 0.001, 0.01, 0.05, 0.1, 0.3, 0.5, 0.7 and 1.0 (please see Fig.\ref{Fig10}).

  \item[(ii)] For $n$ = 1.7, 1.8, 1.9 and 2.0, $Q_{ext}$ versus $\lambda$ can be fitted via a \emph{quartic} regression in the wavelength range $0.11 - 0.40\mu m$ [Fig.\ref{Fig11}(i,iv) and Fig.\ref{Fig12}(i,iv)] and a \emph{quadratic} regression in the wavelength range $0.55 - 0.90\mu m$ [Fig.\ref{Fig11}(ii,v) and Fig.\ref{Fig12}(ii,v)] for $k$ = 0.001, 0.01, 0.05, 0.1 \& 0.3.  Further, $Q_{ext}$ versus $\lambda$ can be fitted via a \emph{quartic} regression in the wavelength range $0.11 - 0.90\mu m$ for higher values of $k$ = 0.5, 0.7 and 1 [Fig.\ref{Fig11}(iii,vi) and Fig.\ref{Fig12}(iii,vi)]. We did not include the plots for $\lambda = 3.4\mu m$.
\end{enumerate}

It is noticed from Figs.\ref{Fig10}, \ref{Fig11} and \ref{Fig12} that $Q_{ext}$ decreases with increase of $\lambda$ when $n \le 1.6$. When $n \ge 1.7$, $Q_{ext}$ initially increases with increase of $\lambda$ and reaches a maximum value, then it starts decreasing if $\lambda$ is increased further. We also observe that $Q_{ext}$ is maximum at $k$ = 0.001 and minimum at $k$ = 1.0 when $\lambda = 0.11\mu m$. But this trend changes at a critical value of wavelength ($\lambda_c$) where exactly opposite nature is noticed. We notice that $\lambda_c$ is (i) 0.16 $\mu m$ at $n = 1.4$, (ii) 0.185 $\mu m$ at $n = 1.5$, (iii) 0.207 $\mu m$ at $n = 1.6$, (iv) 0.22 $\mu m$ at $n = 1.7$, (v) 0.26 $\mu m$ at $n = 1.8$, (vi) 0.26 $\mu m$ at $n = 1.9$ and (vii) 0.30 $\mu m$ at $n = 1.4$. The values of $Q_{ext}$ is maximum at $k$ = 1.0 when $\lambda > \lambda_c$. We do not show any equations and tables in this case.

\subsection{Dependence on the size parameter of monomer ($x$)}
\begin{figure*}
\begin{center}
\includegraphics[]{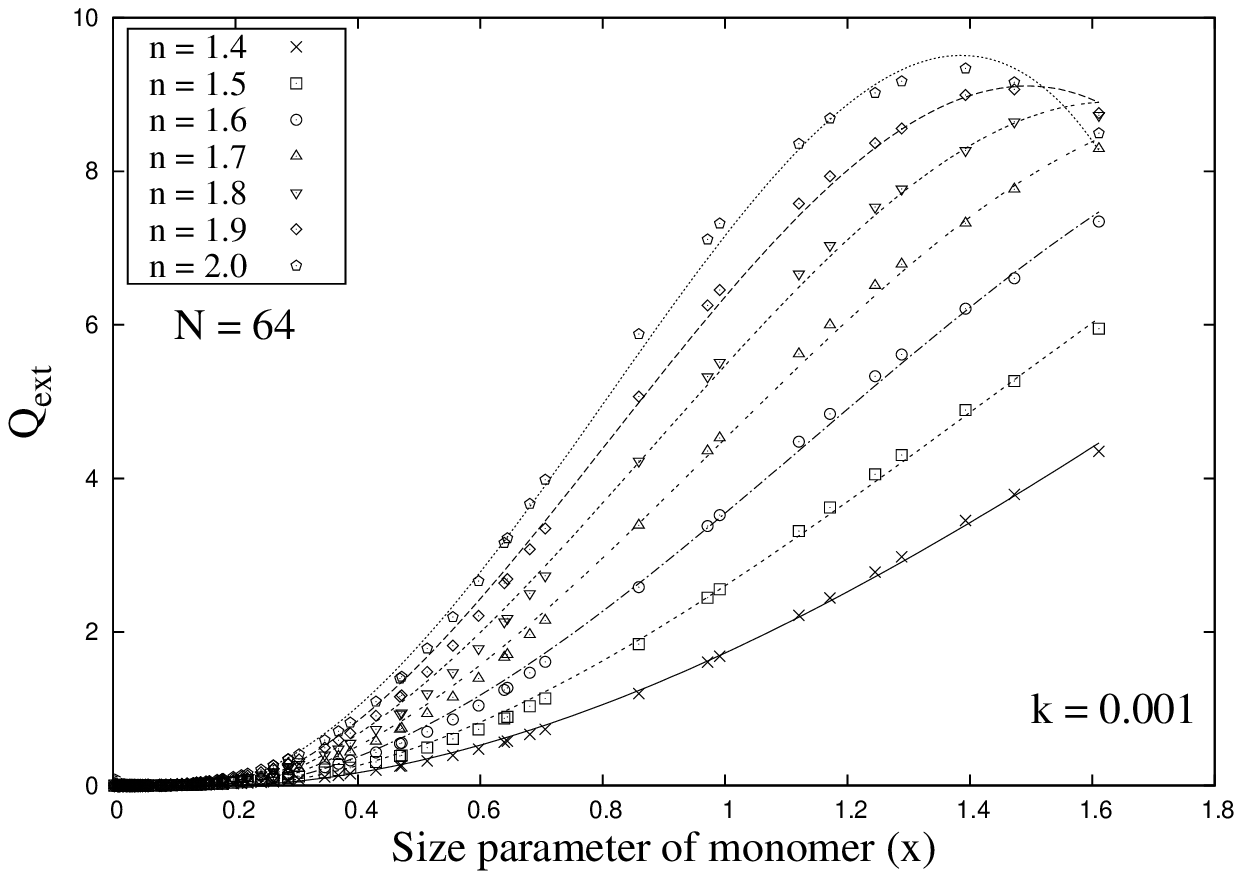}
\caption{Extinction efficiency ($Q_{ext}$) is plotted against the size parameter of monomer ($x = 2\pi a_m/\lambda$, where, $0.01 \le x \le 1.6$) for $N = 64$ and  $k = 0.001$. The best fit curves represent a \emph{cubic regression} for all values of $n$ except 2.0,  where a \emph{quartic regression} (degree 4) is noticed. In all cases, the \emph{coefficient of determination} ($R^2$) $\approx 0.99$. }
\label{Fig13}
\end{center}
\end{figure*}

\begin{figure*}
\begin{center}
\includegraphics[]{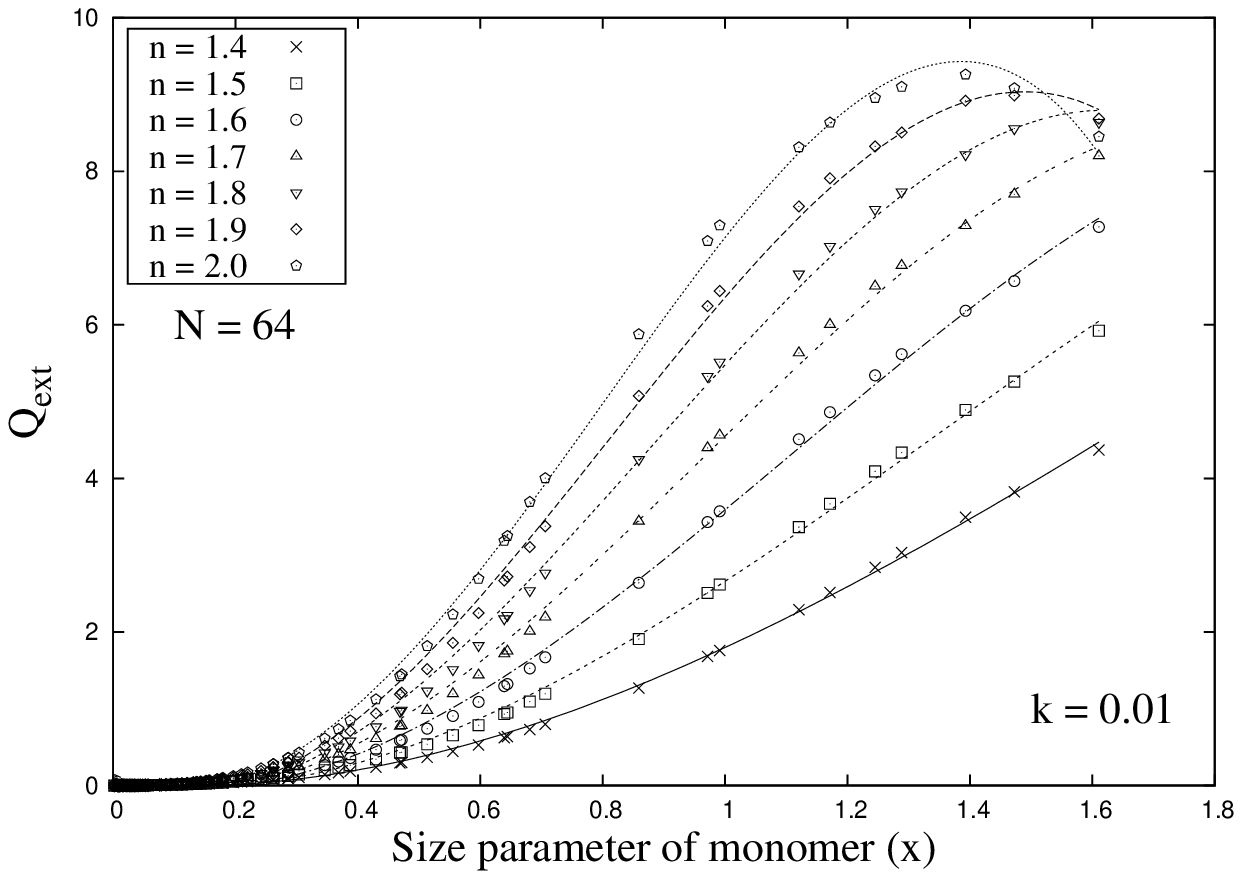}
\caption{Same as Fig.\ref{Fig13} but with $k = 0.01$.}
\label{Fig14}
\end{center}
\end{figure*}

\begin{figure*}
\begin{center}
\includegraphics[]{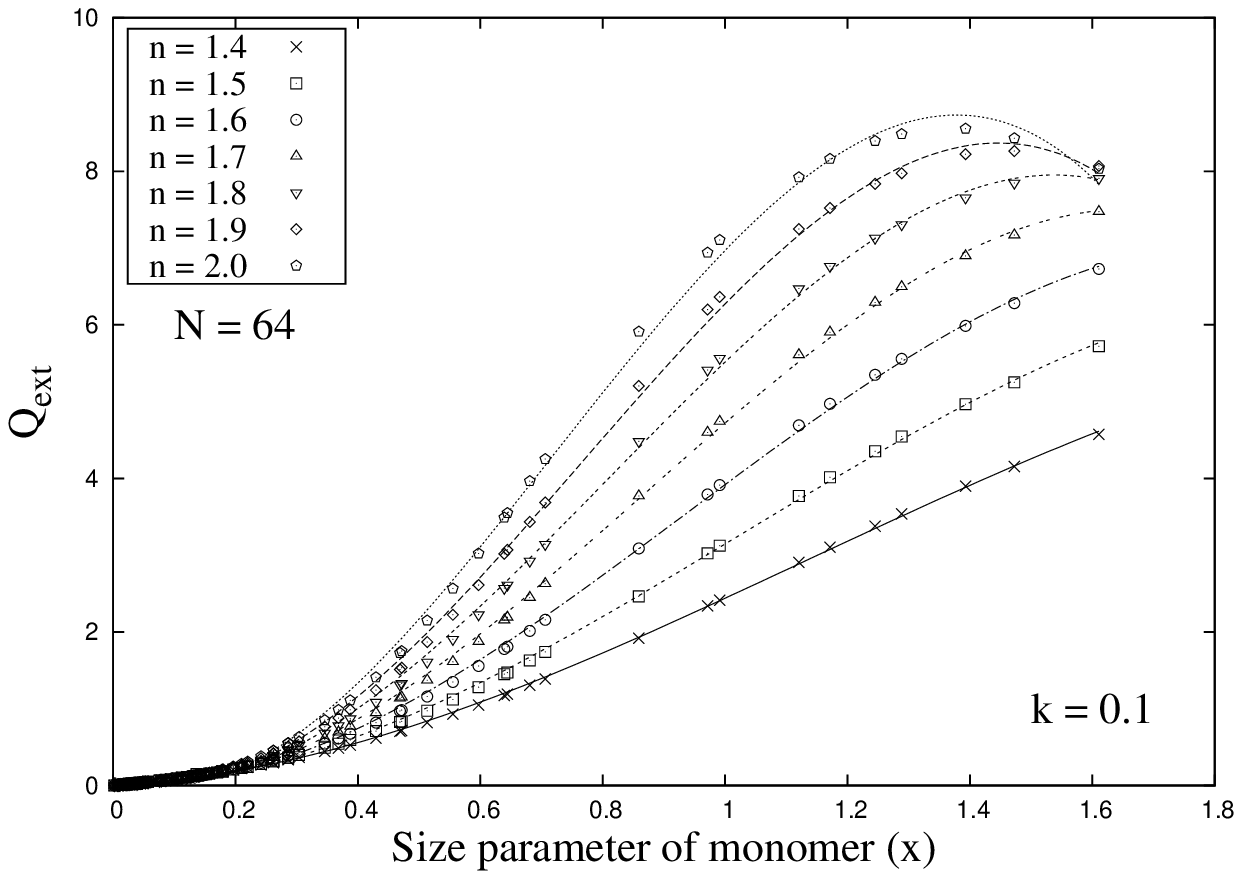}
\caption{Same as Fig.\ref{Fig13} but with $k = 0.1$.}
\label{Fig15}
\end{center}
\end{figure*}

\begin{figure*}
\begin{center}
\includegraphics[]{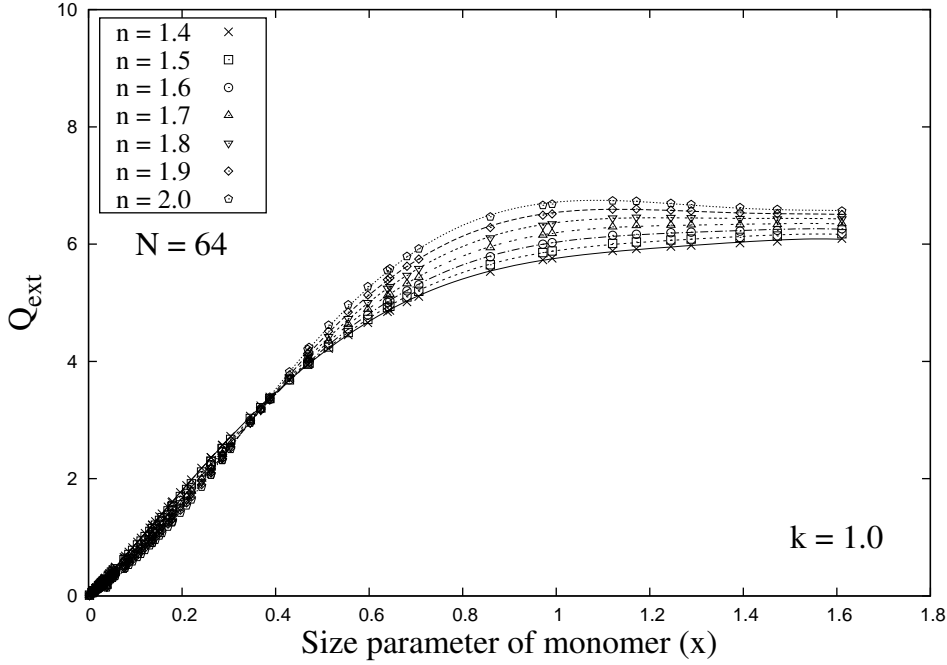}
\caption{$Q_{ext}$ is plotted against $x$ for $N = 64$ and $k = 1.0$. In this case, the best fit curves correspond to a \emph{quintic regression} (degree 5) for all values of $n$. Here, $R^2 \approx 0.99$.}
\label{Fig16}
\end{center}
\end{figure*}

We now study the dependence of $Q_{ext}$ on the size parameter of monomer ($x = 2\pi a_m/\lambda$)  for $n$ = 1.4, 1.5, 1.6, 1.7, 1.8, 1.9, 2.0 and $k$ = 0.001, 0.01, 0.05, 0.1, 0.3, 0.5, 0.7, 1.0.  A wide range of size parameter, $0.01 \le x \le 1.6$ (where, $N = 64$), is considered to investigate the correlation between $Q_{ext}$ and $x$. The results are plotted in Figs. \ref{Fig13}, \ref{Fig14}, \ref{Fig15} and \ref{Fig16}  for $k = 0.001, 0.01, 0.1$ and 1.0, respectively. It can be seen from figures that if $x$ is fixed at any value between 0.01 and 1.6, $Q_{ext}$ increases with increase of $n$. This increase is prominent when $x > 0.2$. The vertical range of $Q_{ext}$ also increases with increase of $n$ from 1.4 to 2.0. This range is maximum  (i) at $x = 1.4$ in case of $k = 0.001$ (where, $Q_{ext}$ = [9.34, 3.45]), (ii) at $x = 1.4$ in case of $k = 0.01$ (where, $Q_{ext}$ = [9.26, 3.50]), (iii) at $x = 1.4$ in case of $k = 0.1$ (where, $Q_{ext}$ = [8.56, 3.90]), and (iv) at $x = 1.12$ in case of $k = 1.0$ (where, $Q_{ext}$(max) = [6.74, 5.88]). The slope of $Q_{ext}$ versus $x$ curve increases with the increase of $n$ from 1.4 to 2.0 which is noted for all values of $k$. It is also interesting to notice that the vertical range of $Q_{ext}$ at $x = 1.6$ decreases with the increase of $k$ and is lowest at $k = 1.0$. Further, $Q_{ext}$ value does not depend much on $n$ for highly absorptive particles ($k = 1.0$) when $x < 0.5$. The variation is also small when $x > 0.5$.

We have found that $Q_{ext}$ and $x$ can be fitted by a \emph{cubic regression} for all values of $n$ except 2.0, in case of $k = 0.001, 0.01$, and 0.1 respectively. The coefficient of determination ($R^2$) for each equation is $\approx 0.99$. The best fit equation is given by
\begin{equation}\tag{4}
    Q_{ext} = \alpha_1 x^3 + \alpha_2 x^2 + \alpha_3 x + \alpha_4,
\end{equation}

where, $\alpha_1$, $\alpha_2$, $\alpha_3$ and $\alpha_4$ are $n$-dependent coefficients of equation (4). The coefficients are presented in Table-\ref{table:5}.

However, the best fit equation in case of $n = 2.0$ corresponds to a \emph{quartic regression} for $k = 0.001, 0.01$, and 0.1, which is given by
\begin{equation}\tag{5}
    Q_{ext} = \beta_1 x^4 + \beta_2 x^3 + \beta_3 x^2 + \beta_4 x + \beta_5,
\end{equation}

where, $\beta_1$, $\beta_2$, $\beta_3$, $\beta_4$ and $\beta_5$ are $n$-dependent coefficients, shown in Table-\ref{table:5}.

The correlation between $Q_{ext}$ and $x$ is found to be \emph{quintic regression} for all values of $n$ at $k = 1.0$. The best fit equation is given by
\begin{equation}\tag{6}
    Q_{ext} = \gamma_1 x^5 + \gamma_2 x^4 + \gamma_3 x^3 + \gamma_4 x^2 + \gamma_5 x + \gamma_6,
\end{equation}
where, $\gamma_1, \gamma_2, \gamma_3, \gamma_4, \gamma_5$, and $\gamma_6$ are $n$-dependent constants of equation (6). The constants are given in Table-\ref{table:5}.

Equations (4), (5) and (6) are very useful in estimating $Q_{ext}$, where one can generate a large data set for $Q_{ext}$ for selected set of $n$, $k$, $a_m$, and $\lambda$.


\begin{table*}
\begin{center}
\caption{All co-efficient of equations (4), (5) and (6).}.
\begin{tabular}{|c|c|c|c|c|c|c|c|}
  \hline
  $k$ & $n$ & $\alpha_1$ & $\alpha_2$ & $\alpha_3$ & $\alpha_4$ &  &  \\
\hline
0.001 & 1.4 & -0.425 & 2.844 & -0.728 & 0.034 &  & \\
      & 1.5 & -1.179 & 5.041 & -1.316 & 0.056 &  & \\
      & 1.6 & -2.239 & 7.681 & -1.976 & 0.082 &  & \\
      & 1.7 & -3.654 &10.775 & -2.712 & 0.109 &  & \\
      & 1.8 & -5.362 &14.173 & -3.475 & 0.136 &  & \\
      & 1.9 & -7.242 &17.624 & -4.175 & 0.158 &  & \\
\hline
0.01  & 1.4 & -0.502 & 2.931 & -0.660 & 0.032 &  & \\
      & 1.5 & -1.248 & 5.085 & -1.231 & 0.055 &  & \\
      & 1.6 & -2.312 & 7.713 & -1.888 & 0.080 &  & \\
      & 1.7 & -3.714 &10.772 & -2.617 & 0.107 &  & \\
      & 1.8 & -5.367 &14.060 & -3.346 & 0.133 &  & \\
      & 1.9 & -7.189 &17.404 & -4.011 & 0.153 &  & \\
\hline
0.1   & 1.4 & -0.891 & 3.035 & 0.273 & 0.023 &  & \\
      & 1.5 & -1.588 & 4.882 &-0.186 & 0.038 &  & \\
      & 1.6 & -2.540 & 7.120 &-0.715 & 0.055 &  & \\
      & 1.7 & -3.726 & 9.652 &-1.276 & 0.072 &  & \\
      & 1.8 & -5.100 &12.358 &-1.829 & 0.087 &  & \\
      & 1.9 & -6.569 &15.046 &-2.302 & 0.097 &  & \\
\hline
\hline
$k$ & $n$ & $\beta_1$ & $\beta_2$ & $\beta_3$ & $\beta_4$ & $\beta_5$ &  \\
\hline
0.001 & 2.0 & -2.164 &-2.689 & 15.031 &-3.120 &0.102 & \\
0.01  & 2.0 & -1.926 &-3.285 & 15.343 &-3.097 &0.103 & \\
0.1   & 2.0 &  0.323 &-8.985 & 18.392 &-2.869 &0.109 & \\
\hline
\hline
$k$ & $n$ & $\gamma_1$ & $\gamma_2$ & $\gamma_3$ & $\gamma_4$ & $\gamma_5$ & $\gamma_6$ \\
\hline
1.0 & 1.4 & -2.709 & 12.371 & -18.540 & 6.182 & 8.466 & -0.009\\
    & 1.5 & -2.973 & 14.016 & -22.167 & 9.438 & 7.576 & -0.003\\
    & 1.6 & -3.253 & 15.774 & -26.024 & 12.846 & 6.686 & 0.005\\
    & 1.7 & -3.540 & 17.603 & -30.058 & 16.378 & 5.800 & 0.014\\
    & 1.8 & -3.861 & 19.625 & -34.440 & 20.118 & 4.901 & 0.024\\
    & 1.9 & -4.222 & 21.861 & -39.192 & 24.073 & 3.988 & 0.035\\
    & 2.0 & -4.625 & 24.319 & -44.317 & 28.233 & 3.061 & 0.047\\
\hline

\end{tabular}
\label{table:5}
\end{center}
\end{table*}

\section{Results from correlation equations}
In the previous sections, we have obtained a set of correlation equations  which can be used to calculate the extinction efficiency of dust aggregates with a wide range of size of aggregates and wavelength of radiation. We first calculate $Q_{ext}$ from relations (2) and (3) for BCCA particles with N = 64 and $a_m = 0.041\mu m$, for selected values of $n, k$ and $\lambda$. The calculated values are compared with the computed values obtained using the Superposition T-matrix code. The results are shown in Table-\ref{table:6} and \ref{table:7}. We also estimate $Q_{ext}$ using relations (4), (5) and (6) for selected values of $x$, $n$ and $k$, and is shown in Table-\ref{table:8}.  It can be seen that the values obtained from computed values match well with the results obtained from correlation equations.

    In general, to model the interstellar extinction, one need to execute the light scattering code with different values
	of $a_m$ (or $a_v$) and wavelength ($\lambda$), which is very time consuming. Using a size distribution for aggregates, it is possible to obtain the average extinction curves for silicate and graphite (and/or amorphous carbon) particles. With a suitable mixing among them, extinction curve against
	different wavelengths can be generated which can be fitted well with observed extinction
	curve. Some good pieces of work on modeling were already done for aggregate particles.
	Some preliminary results on modeling of interstellar extinction using aggregate dust model were already reported by \cite{b1a}.
The present study shows that it is possible to study the extinction properties of interstellar dust aggregates for a given size of the particles and wavelength using relations (4), (5) and (6). The set of  correlation equations can be used to estimate the general extinction $A_\lambda$ using equation (1) for a given size distribution which will help to model the interstellar extinction curve. At this stage, we are not interested in modeling as this study is primarily projected to investigate the dependency of extinction efficiency on size, wavelength and composition of particles. We show how this dependency can be framed with some correlation equations to study the extinction properties of interstellar dust.
	
\cite{b12a} experimentally investigated the
	morphological effects on the extinction band in the infrared region for
	amorphous silica (SiO$_2$) agglomerates. They also compared the measured band profiles
	with calculations for five cluster shapes applying Mie, T-matrix and DDA codes. Our	
	correlations will be also helpful to study the experimental data. But it is
	also important to check the input parameters (size parameter, composition etc.) of the experimental setup before using the correlation equations, because the 	 relations are based on some selected set of parameters.

\begin{table*}
\begin{center}
\caption{$Q_{ext}$ for selected values of $n$ and $k$ from the relation (using equation (2)) and computations (from the simulations) where $a_m = 0.041\mu m$ and $0.11 \le \lambda \le 3.4\mu m$. The difference between correlation equation and computed value is Diff. = $|Q_{ext}$ (corr) $-$ $Q_{ext}$ (comp)$|$}.
\begin{tabular}{|l|l|l|c|c|c|}
  \hline
$\lambda$	&	$n$	&	$k$	&	$Q_{ext}$ (corr)	&	$Q_{ext}$ (comp)	 &	Diff.	\\	
\hline
0.11	&	1.5	&	0.001	&	7.487	&	7.555	&	0.068	\\
	&	1.7	&	0.01	&	8.316	&	8.353	&	0.037	\\
	&	1.9	&	0.1	&	7.729	&	7.683	&	0.046	\\
\hline	
0.20	&	1.5	&	0.001	&	4.368	&	4.305	&	0.063	\\
	&	1.7	&	0.01	&	6.738	&	6.771	&	0.033	\\
	&	1.9	&	0.1	&	7.963	&	7.975	&	0.012	\\
\hline	
0.30	&	1.5	&	0.001	&	1.881	&	1.842	&	0.039	\\
	&	1.7	&	0.01	&	3.431	&	3.441	&	0.010	\\
	&	1.9	&	0.1	&	5.179	&	5.205	&	0.026	\\
\hline	
0.40	&	1.5	&	0.001	&	0.906	&	0.895	&	0.011	\\
	&	1.7	&	0.01	&	1.753	&	1.749	&	0.004	\\
	&	1.9	&	0.1	&	3.057	&	3.070	&	0.013	\\
\hline	
0.55	&	1.5	&	0.001	&	0.390	&	0.388	&	0.002	\\
	&	1.7	&	0.01	&	0.768	&	0.768	&	0.000	\\
	&	1.9	&	0.1	&	1.513	&	1.513	&	0.000	\\
\hline	
0.70	&	1.5	&	0.001	&	0.205	&	0.203	&	0.002	\\
	&	1.7	&	0.01	&	0.406	&	0.406	&	0.000	\\
	&	1.9	&	0.1	&	0.879	&	0.880	&	0.001	\\
\hline	
0.90	&	1.5	&	0.001	&	0.103	&	0.102	&	0.001	\\
	&	1.7	&	0.01	&	0.208	&	0.208	&	0.000	\\
	&	1.9	&	0.1	&	0.508	&	0.509	&	0.001	\\
\hline	
3.4	&	1.5	&	0.001	&	0.002	&	0.002	&	0.001	\\
	&	1.7	&	0.01	&	0.008	&	0.009	&	0.001	\\
	&	1.9	&	0.1	&	0.056	&	0.056	&	0.001	\\
\hline
\end{tabular}
\label{table:6}
\end{center}
\end{table*}

\begin{table*}
\begin{center}
\caption{$Q_{ext}$ for selected values of $n$ and $k$ from the relation (using equation (3)) and computations (from the simulations) where $a_m = 0.041\mu m$ and $0.30 \le \lambda \le 3.4\mu m$ . The difference between correlation equation and computed value is Diff. = $|Q_{ext}$ (corr) $-$ $Q_{ext}$ (comp)$|$}.
\begin{tabular}{|l|l|l|c|c|c|}
  \hline
  $\lambda$	&	$n$	&	$k$	&	$Q_{ext}$ (corr)	&	$Q_{ext}$ (comp)	 &	Diff.	\\	
\hline
0.30	&	1.5	&	0.001	&	1.902	&	1.842	&	0.060	\\
	&	1.7	&	0.01	&	3.438	&	3.441	&	0.003	\\
	&	1.9	&	0.1	&	5.180	&	5.205	&	0.025	\\
\hline	
0.40	&	1.5	&	0.001	&	0.918	&	0.895	&	0.023	\\
	&	1.7	&	0.01	&	1.760	&	1.749	&	0.011	\\
	&	1.9	&	0.1	&	3.049	&	3.070	&	0.021	\\
\hline	
0.55	&	1.5	&	0.001	&	0.394	&	0.388	&	0.006	\\
	&	1.7	&	0.01	&	0.770	&	0.768	&	0.002	\\
	&	1.9	&	0.1	&	1.513	&	1.513	&	0.000	\\
\hline	
0.70	&	1.5	&	0.001	&	0.205	&	0.203	&	0.002	\\
	&	1.7	&	0.01	&	0.407	&	0.406	&	0.001	\\
	&	1.9	&	0.1	&	0.879	&	0.880	&	0.001	\\
\hline	
0.90	&	1.5	&	0.001	&	0.102	&	0.102	&	0.000	\\
	&	1.7	&	0.01	&	0.208	&	0.208	&	0.000	\\
	&	1.9	&	0.1	&	0.509	&	0.509	&	0.000	\\
\hline	
3.4	&	1.5	&	0.001	&	0.002	&	0.002	&	0.000	\\
	&	1.7	&	0.01	&	0.008	&	0.009	&	0.001	\\
	&	1.9	&	0.1	&	0.056	&	0.056	&	0.000	\\
\hline
\end{tabular}
\label{table:7}
\end{center}
\end{table*}

\begin{table*}
\begin{center}
\caption{$Q_{ext}$ for selected values of $n$, $k$ and $x$ from the relation (using equations (4), (5) and (6)) and computations (from the simulations). The difference between correlation equation and computed value is Diff. = $|Q_{ext}$ (corr) $-$ $Q_{ext}$ (comp)$|$}.
\begin{tabular}{|l|l|l|c|c|c|}
  \hline
~~$x$	&	$n$	&	$k$	&	$Q_{ext}$ (corr)	&	$Q_{ext}$ (comp)	 &	 Diff.	\\	
\hline
0.286	&	1.5	&	0.001	&	0.064	&	0.092	&	0.028	\\
0.286	&	1.7	&	0.01	&	0.152	&	0.207	&	0.055	\\
0.286	&	1.9	&	0.1	&	0.514	&	0.506	&	0.008	\\
\hline											
0.468	&	1.5	&	0.001	&	0.424	&	0.388	&	0.036	\\
0.468	&	1.7	&	0.01	&	0.862	&	0.768	&	0.094	\\
0.468	&	2.0	&	1.0	&	4.183	&	4.205	&	0.022	\\
\hline											
0.706	&	1.5	&	0.001	&	1.225	&	1.136	&	0.089	\\
0.706	&	1.9	&	0.1	&	3.660	&	3.688	&	0.028	\\
0.706	&	2.0	&	0.1	&	4.169	&	4.252	&	0.083	\\
\hline											
0.972	&	1.5	&	0.001	&	2.459	&	2.446	&	0.013	\\
0.972	&	1.7	&	0.01	&	4.333	&	4.397	&	0.064	\\
0.972	&	2.0	&	1.0	&	6.693	&	6.664	&	0.029	\\
\hline											
1.171	&	1.5	&	0.001	&	3.534	&	3.622	&	0.088	\\
1.171	&	1.9	&	0.1	&	7.485	&	7.522	&	0.037	\\
1.171	&	2.0	&	1.0	&	6.728	&	6.733	&	0.005	\\
\hline											
1.289	&	1.5	&	0.001	&	4.208	&	4.305	&	0.097	\\
1.289	&	1.7	&	0.01	&	6.674	&	6.771	&	0.097	\\
1.289	&	1.9	&	0.1	&	8.058	&	7.975	&	0.083	\\
\hline											
1.473	&	1.5	&	0.001	&	5.287	&	5.268	&	0.019	\\
1.473	&	1.7	&	0.01	&	7.754	&	7.702	&	0.052	\\
1.473	&	1.9	&	0.1	&	8.357	&	8.264	&	0.093	\\
\hline
\end{tabular}
\label{table:8}
\end{center}
\end{table*}

\section{Summary}
\begin{enumerate}
  \item  We have first studied the dependency of $Q_{ext}$ on size of the aggregates ($a_m$) to investigate the correlations between $Q_{ext}$ and complex refractive indices ($n, k$) at a particular size. Computations are performed at four different sizes ($a_v = 0.004, 0.068, 0.16$ and 0.26 $\mu m$).

If $k$ is fixed at any value between 0.001 and 1.0, $Q_{ext}$ increases with increase of $n$ from 1.4 to 2.0. $Q_{ext}$ and $n$ are correlated  via \emph{linear} regression when the cluster size is small whereas the correlation is \emph{quadratic} at moderate and higher sizes of cluster. This feature is observed at all wavelengths (UV to optical to infrared). We have also found that the variation of $Q_{ext}$ with $n$ is very small when $\lambda$ is high.

We have observed that  $Q_{ext}$ and $k$ are correlated  via polynomial regression equation (of degree 2,3 or 4) where the degree of equation depends on the cluster size, real  part of the refractive index of the particles ($n$) and wavelength ($\lambda$) of incident radiation. At $a_v = 0.16 \mu m$, $Q_{ext}$ and $k$ is found to be correlated with a polynomial regression equation (of degree 2 or 3) when $\lambda$ is between 0.11 $\mu m$ and 0.26 $\mu m$. However, when $\lambda > 0.26\mu m$,  we have found that the correlation between them is \emph{quadratic} for all values of $n$. The vertical range of $Q_{ext}$ in the plot also decreases when $k$ increases. This range is maximum at $k = 0.001$  and minimum at $k = 1.0$. If we include results for four different cluster sizes, we can summarize that the correlation of $Q_{ext}$ and $k$ depends mainly on cluster size. The correlation is linear for small size and quadratic/cubic/quartic for moderate and higher sizes.

\item We study the dependence of $Q_{ext}$ on $\lambda$ for $a_v = 0.16\mu m$. $Q_{ext}$ decreases with increase of $\lambda$ when $n \le 1.6$. When $n \ge 1.7$, $Q_{ext}$ initially increases with increase of $\lambda$ and reaches a maximum value, then it starts decreasing if $\lambda$ is increased further. For $n$ = 1.4, 1.5 and 1.6, $Q_{ext}$ versus $\lambda$ can be fitted via a \emph{quartic} regression for all values of $k$. For other values of $n$, the correlation is polynomial regression where the degree of equation depends on the value of $n, k$ and $\lambda$.

\item We have found that $Q_{ext}$ and $x$ are correlated via a polynomial regression (of degree 3,4 or 5) for all values of $n$. The degree of regression is found to be $n$ and $k$-dependent.

   \item The correlation equations can be used to model interstellar extinction for dust aggregates in a wide range of size of the aggregates, wavelengths and complex refractive indices.
\end{enumerate}


\section{Acknowledgement}
We acknowledge Daniel Mackowski and Michael Mishchenko, who made their Multi-sphere T-matrix (MSTM) code  publicly available. We also acknowledge Prithish Halder for help on the execution of JaSTA-2 software package. The reviewer of this paper is highly acknowledged for useful comments and suggestions.

\end{document}